\documentclass[nofootinbib,aps,pra,a4paper,preprintnumbers,showpacs,showkeys]{revtex4-1}

\usepackage{amssymb}
\usepackage{amsmath}
\usepackage{graphicx}
\usepackage{dcolumn}
\usepackage{bm}
\usepackage{epsfig}
\usepackage{hyperref}

\newcommand\beq{\begin{eqnarray}}
\newcommand\eeq{\end{eqnarray}} 
 
\newcommand\eq[1]{Eq. (\ref{eq:#1})} 

\newcommand\figwidth{.25\textwidth}
\newcommand\Eq[1]{Eq.~\ref{eq:#1}}
\newcommand\Fig[1]{Fig.~\ref{fig:#1}}

\newcommand\bfn{\mathbf n}

\newcommand\bfm{\mathbf m}

\newcommand\bfe{\mathbf e}

\newcommand\calA{\mathcal A}
\newcommand\calB{\mathcal B}
\newcommand\calS{\mathcal S}
\newcommand\calF{\mathcal F}
\newcommand\calC{\mathcal C}
\newcommand\calD{\mathcal D}

\newcommand\calE{\mathcal E}
\newcommand\calL{\mathcal L}

\newcommand\calV{\mathcal V}
\newcommand{\Tr}{{\rm Tr\,}}
\newcommand{\diag}{{\rm diag\,}}

\begin{document}

\preprint{INT-PUB-12-021, RIKEN-QHP-21}

\title{Lattice theory for nonrelativistic fermions in one spatial dimension}

\author{Michael G. Endres}
\email{endres@riken.jp}

\affiliation{Theoretical Research Division, RIKEN Nishina Center, Wako, Saitama 351-0198, Japan}


\date{\today}
 
\begin{abstract}
I derive a loop representation for the canonical and grand-canonical partition functions for an interacting four-component Fermi gas in one spatial dimension and an arbitrary external potential.
The representation is free of the ``sign problem'' irrespective of population imbalance, mass imbalance, and to a degree, sign of the interaction strength.
This property is in sharp contrast with the analogous three-dimensional two-component interacting Fermi gas, which exhibits a sign problem in the case of unequal masses, chemical potentials, and repulsive interactions.
The one-dimensional system is believed to exhibit many phenomena in common with its three-dimensional counterpart, including an analog of the BCS-BEC crossover, and nonperturbative universal few- and many-body physics at scattering lengths much larger than the range of interaction, making the theory an interesting candidate for numerical study.
Positivity of the probability measure for the partition function allows for a mean-field treatment of the model;
here, I present such an analysis for the interacting Fermi gas in the $SU(4)$ (unpolarized, mass-symmetric) limit, and demonstrate that there exists a phase in which a continuum limit may be defined.
\end{abstract}

\maketitle

\section{Introduction}

In recent years, nonrelativistic Fermi gases have become the focus of intense study, fueled primarily by significant advances in ultra-cold atom experiments \cite{RevModPhys.80.1215}.
Of primary interest has been the BCS-BEC crossover regime for a dilute two-component Fermi gas with an attractive short-range two-body interaction in three spatial dimensions.
In the limit that the scattering length is much larger than the range of interactions, this system, often referred to as a ``unitary Fermi gas,'' becomes universal in the sense that the long distance physics is independent of the details of the inter-particle potential.
In the unitary limit, the Fermi gas is characterized by a single dimensionful scale, namely the density of the system, and as a consequence, the energy of the interacting gas is proportional to the free-gas energy, and the pairing gap for fermion excitations is proportional to the Fermi energy.
This conformal and scale invariant system is not only experimentally realizable using ultra-cold atoms tuned to a Feshbach resonance, but also provides an idealized theoretical description of dilute nuclear matter \cite{baker2000mbx}.

The unitary Fermi gas is a challenging system to explore theoretically due to the strongly coupled nature of the interaction, as well as a lack of scales in the problem.
Although qualitatively much can be said about unitary fermions, a nonperturbative treatment is essential for achieving a complete quantitative understanding of the system.
Through the use of numerical techniques, considerable headway has been made in exploring the properties of the few- and many-body system both at zero and at finite temperature.
Monte Carlo techniques have been particularly successful in studies of the unpolarized unitary Fermi gas, in which each spin component has the same mass (for recent studies, see e.g., \cite{PhysRevA.83.041601,PhysRevLett.106.235303,PhysRevA.84.061602,PhysRevLett.106.205302}), and the slightly polarized gas, in which a single fermion has been removed \cite{PhysRevLett.100.150403}.

Introducing a population imbalance in the system by means of unequal chemical potentials for each species in the grand-canonical ensemble, a mass imbalance, or repulsive interaction results in a sign or signal/noise problem, making numerical calculations which rely on Monte Carlo simulations a formidable challenge.
Despite the considerable attention such systems have received \cite{2012arXiv1205.0568G}, from a theoretical standpoint less is known quantitatively about the phase diagram in each of these scenarios.
Recently, progress has been made from the experimental side, however \cite{Jo18092009,2009IJMPB..23.3195L}.
The mass-imbalanced unitary Fermi gas is particularly interesting because the scale-invariance of the system is expected to break down to a discrete scale invariance at a critical mass ratio, leading to Effimov physics \cite{Efimov1973157,Braaten2006259}.

The realization of optical lattices in recent years make experimental exploration of Fermi gases in lower and mixed spatial dimensions a feasible and attractive alternative \cite{RevModPhys.80.885}.
Theoretical investigations suggest that such systems can also exhibit scale invariance and universal properties, much in the same way as their two-component, three-dimensional Fermi gas counterpart \cite{PhysRevLett.101.170401}.
One specific example, which will be the focus of this paper, is the four-component Fermi gas in one spatial dimension with an attractive short-range four-body interaction.
It was shown that this system is also stable and universal in the sense that its properties are independent of the confining potential when tuned to a four-body resonance \cite{Nishida:2009pg}. 

The relevant dimensionful scales which describe the one-dimensional Fermi gas are the Fermi momentum $p_F = \pi n/4$, where $n$ is the number density of the system, the Fermi energy $E_F = p_F^2/(2m)$, where $m$ is the mass of the fermions, and the scattering length $a$, which characterizes the long-distance properties of the four-body interaction.
As with the universal Fermi gas in three dimensions, the properties of this model are smoothly controlled by a single dimensionless parameter given by $\eta = (p_F a)^{-1}$ \cite{Nishida:2009pg}.
In the regime $\eta \ll -1$, the system forms a one-dimensional analog of a BCS superfluid, with three gaps for fermion excitations, each of which are exponentially small in $\eta$.
On the other hand, in the regime $\eta \gg 1$, the many-body system forms a dilute Bose-Einstein condensate of deeply bound tetramers with largely gapped fermion excitations proportional to $1/(m a^2)$.
Each of these regimes may be understood within the framework of effective field theories; in the former case, the low-energy constants entering the effective theory can be matched onto the underlying microscopic theory perturbatively in $|\eta|^{-1}$, but in the latter case, the matching must be done nonperturbatively by solving the eight-body problem \cite{Nishida:2009pg}.

In the cross-over region corresponding to $|\eta| \ll 1$, there remains only one dimensionful parameter to describe the system, namely the density.
Consequently, by dimensional analysis, the ground state energy density must be proportional to the free-gas energy density, $\calE_{FG} = n E_F/3$.
Similarly, the excitation gaps are in proportion to the Fermi energy. 
As is the case for the two-component three-dimensional Fermi gas, the four-component one-dimensional Fermi gas satisfies universal relations (i.e., Tan relations) involving a recently introduced parameter, known as the ``contact density'' \cite{2008AnPhy.323.2952T,2008AnPhy.323.2971T,2008AnPhy.323.2987T,2008PhRvA..78e3606B,PhysRevLett.100.205301}, which is related to the fall-off of the tail of the momentum distribution of the Fermi gas.
In addition, as a result of an operator-state correspondence \cite{PhysRevD.76.086004}, an exact relationship exists between the energies of the Fermi gas at infinite scattering length when confined to an external harmonic trap, and the scaling dimension of primary operators in free-space, thus motivating nonperturbative studies of the trapped few-body system.

Because the one-dimensional Fermi gas is nonperturbative in the cross-over regime, numerical simulations are crucial for obtaining a quantitative understanding of the few- and many-body physics of this model.
In this paper, I show that an effective theory description of the interacting four-component Fermi gas in one spatial dimension can be formulated on a lattice in a way that avoids the sign problem.
Surprisingly, the sign problem remains absent in the new path-integral representation for unequal masses and chemical potentials, in the presence of an arbitrary external potential, and even to an extent, irrespective of the sign of the four-body interaction strength.\footnote{For a review of various other one-dimensional models for which the sign problem has been solved, see \cite{doi:10.1080/0001873021000049195} and references within.}
The canonical and grand-canonical partition functions are formulated in terms of a path-integral over self-avoiding non-intersecting fermion loops, and are suitable for numerical study using a simple loop-updating scheme, or more sophisticated algorithms.
I derive expressions for both of these partition functions, as well as for expectation values of the kinetic, potential and interaction energies, charges and their associated susceptibilities, and $N$-point correlation functions involving local density operators.
Finally, I explore the number density phase diagram of the lattice theory at zero temperature, and demonstrate that, at least within mean-field theory, a continuum limit exists.

\section{Lattice theory}

In the universal regime, the thermodynamics of a dilute one-dimensional four-component Fermi gas may be reliably described by the continuum effective field theory:
\begin{eqnarray}
S_{cont}(\mu) =\int d^2x \, \left[ \psi^\dagger \left( \partial_\tau - \frac{\nabla^2}{2m} - \mu \right) \psi - g (\psi^\dagger \psi)^4 \right]\ ,
\label{eq:continuum_action}
\end{eqnarray}
defined in Euclidean spacetime.
Here, $\psi = (\psi_a, \psi_b, \psi_c, \psi_d)$ is a four-component Grassmann-valued spinor, and $m = \diag (m_a, m_b, m_c, m_d)$ and  $\mu = \diag (\mu_a, \mu_b, \mu_c, \mu_d)$ are diagonal mass and chemical potential matrices, respectively.
Fermion interactions are achieved via a four-fermion contact potential, with strength $g$ tuned to reproduce the physical four-body scattering length.
I discretize \Eq{continuum_action} on an $N_\tau \times N_s$ lattice $\Lambda$, imposing anti-periodic boundary conditions in time and open boundary conditions in space \footnote{Note that any choice of spatial boundary conditions are acceptable, however, the choice of open spatial boundary conditions simplifies the forthcoming analysis considerably.}.
The lattice sites are labeled by the integer coordinate pair $\bfn = (n_\tau, n_s)$ with $n_\tau\in [0,N_\tau-1]$ and $n_s \in [0,N_s-1]$.
The physical length of the lattice is given by $L=b_s N_s$ and the inverse temperature by $\beta = b_\tau N_\tau$, where $b_s$ and $b_\tau$ are the spatial and temporal lattice spacings, respectively.
Following the approach of \cite{Chen:2003vy}, the lattice-discretized action is obtained by replacing the continuous fields by their discrete counterpart:  $\psi(x) \to \psi_\bfn$, the integral in \Eq{continuum_action} with a discrete sum over lattice sites: $\int d^2x \to b_\tau b_s\sum_{\bfn\in\Lambda}$, and identifying the terms in the action as follows:
\begin{eqnarray}
(\partial_\tau \psi - \mu\psi)_\bfn &=  & \frac{1}{b_\tau} \left( \psi_\bfn - e^{b_\tau \mu} \psi_{\bfn-{b_\tau \bfe_0}} \right) \ ,\cr
(-\nabla^2 \psi)_\bfn               &=  & \frac{1}{b_s^2} \left( 2\psi_\bfn -\psi_{\bfn+{b_s \bfe_1}} - \psi_{\bfn-{b_s \bfe_1}} \right) \ ,\cr
(\psi^\dagger \psi)_\bfn            &=  & \psi_\bfn^\dagger e^{b_\tau \mu} \psi_{\bfn-{b_\tau \bfe_0}} \ ,
\end{eqnarray}
where $\bfe_j$ are unit vectors in the time ($j=0$) and space ($j=1$) directions of the lattice.
The continuum limits in space and time correspond to taking all scales with dimensions of length much larger than $b_s$ and all scales with dimensions of time much larger than $b_\tau$.
For simplicity, I will work in ``lattice units'' where the temporal and spatial lattice spacings (along with $\hbar$) are set to unity, and where all other dimensionful quantities are then measured in appropriate units of the lattice spacings.

Note that \Eq{continuum_action} is a highly tuned action in the sense that two- and three-body interactions are absent.
Radiative corrections giving rise to such operators may be avoided without fine-tuning of the lattice action through the use of a point-split four-body interaction, which disallows the formation of fermion loops on a fixed time slice.
At infinite volume and for an attractive interaction, the binding energy $E$ of the four-fermion system on the lattice may be related to the coupling by solutions to the integral equation:
\begin{eqnarray}
\frac{1}{2\pi g} =  \int_{-\pi}^\pi \left( \prod_\sigma{\frac{d p_\sigma}{2\pi}} \right)  \frac{\delta(\sum_\sigma p_\sigma) }{e^{-E} \prod_\sigma \xi_{p_\sigma}(m_\sigma) -1}\ ,
\label{eq:four_body}
\end{eqnarray}
where $\xi_p(m) = 1 + \Delta_p/m$, and $\Delta_p = 2 \sin^2(p/2)$.
This result is obtained by diagonalizing the four-fermion transfer matrix associated with \Eq{continuum_action} on the lattice at zero chemical potential, following the approach of \cite{Endres:2012cw}. 

In the $SU(4)$ symmetric limit ($m_\sigma = m$ for all $\sigma$), the four-fermion coupling may also be related to the four-particle scattering length by evaluating the inverse scattering amplitude $\calA^{-1}(p)$ for four-fermion scattering at vanishing external momentum $p = \sqrt{m E}$, and equating the result to $m/(4\pi a) $.
This calculation may be performed on the lattice following a prescription analogous to \cite{1996NuPhB.478..629K,Chen:2003vy}, by summing the Feynman diagrams for four-particle scattering in a vacuum as a geometric series.
Combining that result with \Eq{four_body}, yields the relation:
\begin{eqnarray}
-\frac{m}{4\pi a} = \frac{1}{g} - \frac{1}{g_{c}}  \ .
\end{eqnarray}
The unitary limit (i.e., the limit $a=\infty$) is obtained by tuning the coupling $g$ to some critical value $g_c$, corresponding to a zero-energy four-body bound state (i.e., a solution to \Eq{four_body} with $E=0$).
At finite but asymptotically large positive scattering lengths, the binding energy of the system is given by $E\sim -1/(2 m a^2)$.
Note that negative values of the coupling are required when $m/(4\pi a)$ exceeds $1/g_c$, and in conventional path-integral representations this usually implies a sign problem.
As will be demonstrated in the proceeding sections, however, in the loop representation the sign problem is avoided for an expanded parameter regime governed by $g > -1$.

\section{Loop representation}

The grand-canonical partition function for the lattice theory is given by
\begin{eqnarray}
Z(\mu) = \int [d\psi^\dagger][d\psi] e^{ - S(\mu) } \ ,
\label{eq:partition_func}
\end{eqnarray}
where
\begin{eqnarray}
-S(\mu) = \sum_{\bfn\in\Lambda} \left[
              - \psi^\dagger_\bfn \left(1+\frac{1}{m} \right) \psi_\bfn
              + \psi^\dagger_\bfn e^{\mu} \psi_{\bfn-\bfe_0}
              + \psi^\dagger_\bfn \frac{1}{2m} \left( \psi_{\bfn+\bfe_1} + \psi_{\bfn-\bfe_1} \right)
              + g \left( \psi^\dagger_\bfn e^\mu \psi_{\bfn-\bfe_0} \right)^4
               \right]\ .
\end{eqnarray}
One may express \Eq{partition_func} in terms of a hopping parameter expansion \cite{itzykson:book}, by expanding $e^{- S(\mu)}$ in a power-series, followed by explicit integration over fermion fields.
Note that at finite space-time volume, the expansion always truncates after a finite number of terms and is therefore well-defined, and furthermore converges (although the number of nonvanishing contributions will grow exponentially with volume).
The contributions to the hopping parameter expansion from the various terms in the lattice action are given as follows:
\begin{eqnarray}
\psi_\bfn \psi^\dagger_\bfn                  \sim -\left( 1+\frac{1}{m} \right) \ ,\quad
\psi_\bfn \psi^\dagger_{\bfn+{\bfe_0}}       \sim e^\mu                         \ ,\quad
\psi_\bfn \psi^\dagger_{\bfn\pm{\bfe_1}}     \sim \frac{1}{2m}                  \ ,\quad
(\psi^\dagger_\bfn \psi_{\bfn-{\bfe_0}})^4   \sim g e^{\Tr \mu }                            \ .
\label{eq:weights}
\end{eqnarray}
Upon integration over fermions, the surviving terms in the expansion are those in which all lattice sites $\bfn\in\Lambda$ are occupied by a single power of $\psi_\sigma^\dagger \psi_\sigma$ for each species $\sigma = \{a,b,c,d\}$; multiple insertions of $\psi_\sigma^\dagger \psi_\sigma$ at the same site for a given species must vanish due to the anti-commuting nature of Grassmann numbers.

Nonvanishing configurations in the hopping parameter expansion may be specified diagrammatically in a unique way as a configuration of monomer and tetramer bonds associated with insertions (either zero or one) of either the nearest neighbor or four-fermion interaction at various locations of the lattice; for a given configuration, all remaining unoccupied lattice sites are saturated by insertions of the on-site interaction.
Non-vanishing configurations under fermion integration satisfy the constraint that monomers form closed loops; for all such closed loop configurations, a tetramer may occupy time-like bonds in lieu of monomers, provided the bond was occupied by a monomer of every species.
The contribution to the partition function from such configurations is simply the product of weights associated with each insertion of on-site, nearest neighbor and four-fermion operators, provided in \Eq{weights}.

Note that the hopping parameter expansion exhibits several convenient properties in one spatial dimension:
\begin{itemize}
\item fermion loops formed by monomers of the same species are self-avoiding due to Pauli exclusion with one exception: fermion loops may form on a fixed time-slice provided they only involve a single hop forward and back (or vice-versa); for every such loop, the configuration is weighted by a negative sign;
\item fermion loops may form by wrapping around the time direction and yield contributions to the partition function proportional to powers of $e^{N_\tau \mu}$; fermion loops may not wrap around the lattice in space, however, as a result of employing open spatial boundary conditions;
\item a fermion loop cannot wrap more than once around the lattice in time since closure of the loop would require the fermion path to cross itself.
Multiple fermion loops are permitted provided each loop wraps around the time direction only once;
the sign arising from every fermion loop of this kind cancels with the sign associated with anti-periodic temporal boundary conditions;
\item the number of sites that are saturated by insertions of the on-site interaction is even for every configuration provided $N_\tau$ is even, therefore the sign associated with the weight for on-site interactions may be ignored.
\end{itemize}

\begin{figure}
\includegraphics[width=\figwidth]{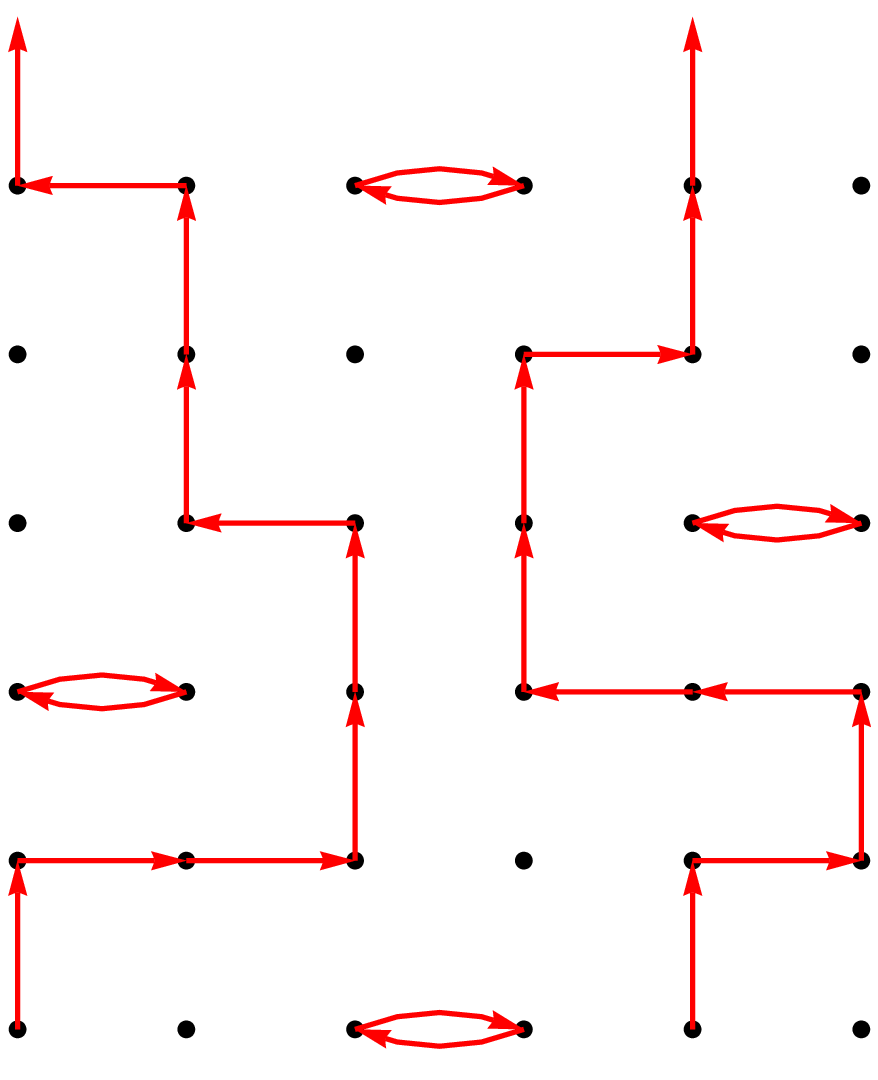}
\hspace{0.5in}
\includegraphics[width=\figwidth]{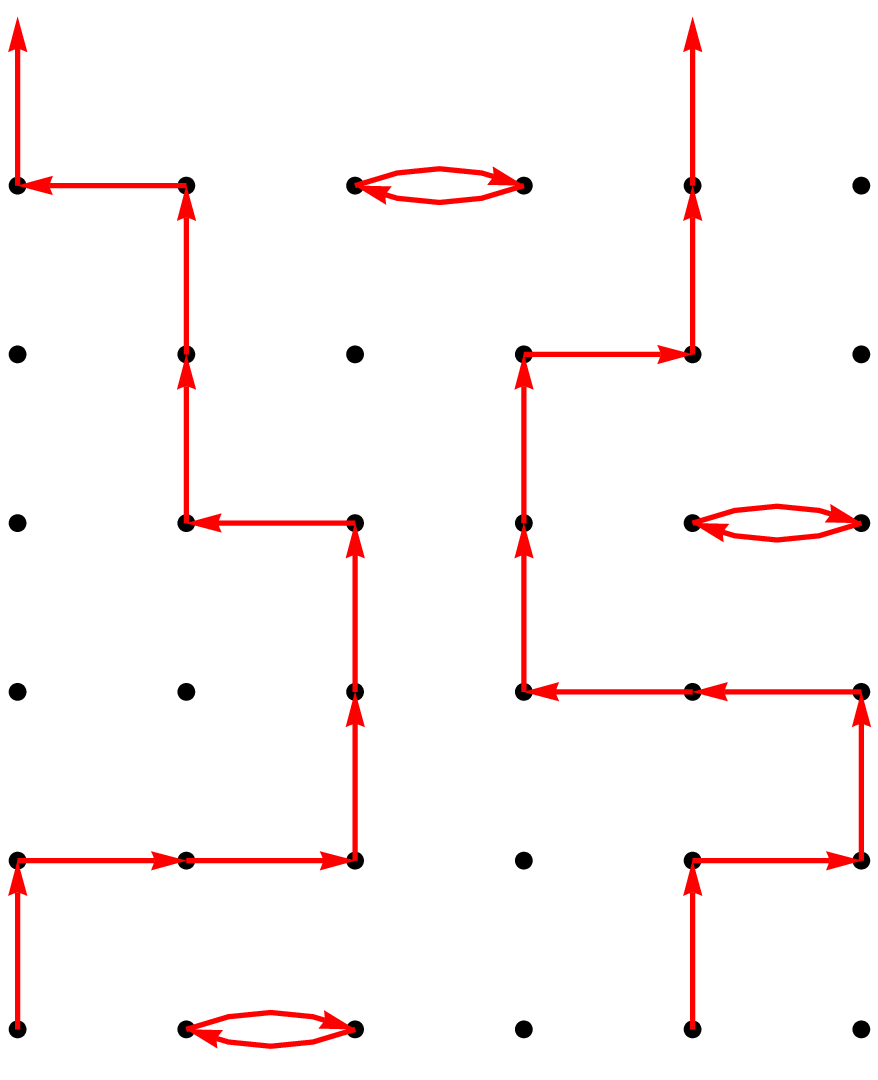}
\hspace{0.5in}
\includegraphics[width=\figwidth]{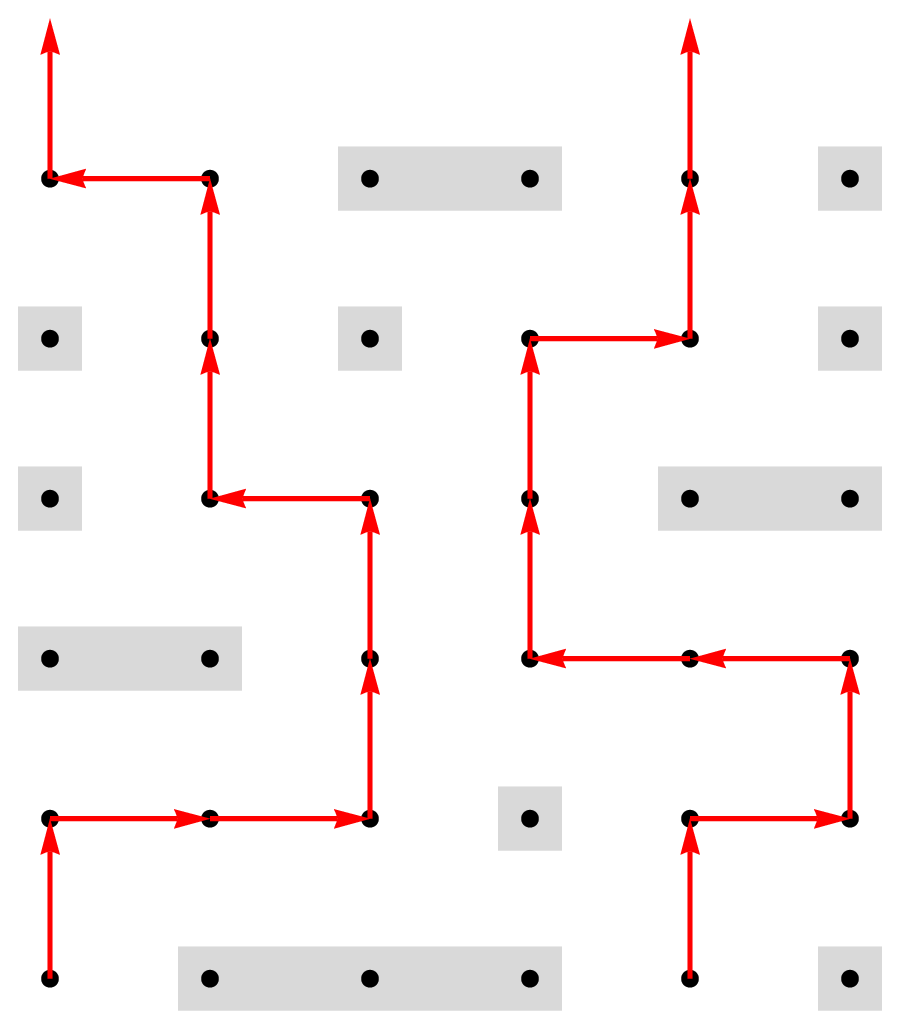}
\caption{%
\label{fig:config}%
Example of three class-equivalent closed-loop configurations belonging to the set $C$ on an $N_\tau\times N_s = 6\times 6$ lattice.
The left-most configuration has $\calS = 8$, $\calB_s = 16$, $\calB_\tau = 12$ and $\calF= 4$, whereas the center configuration has $\calS = 10$, $\calB_s=14$, $\calB_\tau = 12$ and $\calF= 3$.
The right-most configuration belongs to the subset $C^*$ of bubble-free configurations and has $\calS = 8$, $\calB_s = 8$, $\calB_\tau = 12$ and $\calF= 0$;
this configuration has eleven domains $d\in\calD(c)$ (shaded) of which seven have $\calL(d) = 1$, three have $\calL(d)=2$ and one has $\calL(d) = 3$.
}
\end{figure}

With these considerations in mind, the partition function may be explicitly expressed as
\begin{eqnarray}
Z(\mu) = \sum_{ \{ c_\sigma \} \in C } \left[  \prod_\sigma  \left( 1+ \frac{1}{m_\sigma} \right)^{\calS(c_\sigma)} \left(\frac{1}{2m_\sigma} \right)^{\calB_s(c_\sigma)} \left(-1\right)^{\calF(c_\sigma)} e^{\mu_\sigma \calB_\tau(c_\sigma)} \right] \left(1 + g \right)^{\calB_\tau( \cap_\sigma c_\sigma) } \ ,
\label{eq:loops_with_bubbles}
\end{eqnarray}
where $C$ is the set of all possible configurations involving self-avoiding, closed-loops which wrap around the time direction and space-like single-hop fermion bubbles.
An example of two such configurations for a single species is shown in \Fig{config} (left, center).
In \Eq{loops_with_bubbles}, $\calS(c)$ represents the total number of sites left unvisited by loops associated with $c$, $\calB_s(c)$ and $\calB_\tau(c)$ are the total number of space-like and time-like bonds lying on the loops associated with $c$, and $\calF(c)$ is the total number of space-like fermion bubble loops associated with $c$.
Note that these quantities satisfy the relationship $N_\tau N_s = \calB_\tau + \calB_s + \calS$; since $\calB_\tau \propto N_\tau$ and  $\calB_s$ is always even, if $N_\tau$ is even then so is $\calS$.
Thus, I am justified in dropping the negative sign associated with the weight of the on-site interaction.

Note that the quantity $\calB_\tau(c) / N_\tau$  is an integer, and may be regarded as a fermion charge associated with configuration $c$.
Particularly, this quantity specifies the winding number of the configuration.
Viewing a particular configuration $c$ as a set of occupied bonds, $\calB_\tau( \cap_\sigma c_\sigma )$ is defined as the number of time-like bonds in the intersection $\cap_\sigma c_\sigma$.
Such intersections may be associated with either four insertions of a monomer bond (one of each species) with a weight of unity or a single insertion of a tetramer bond with a weight of $g$, up to a common factor of $e^{\Tr\mu}$ which is accounted for separately.
For every fixed  set of configurations $\{c_\sigma\}$, I have summed over both possibilities, leading to an overall  weight of $1+g$ associated with the intersection on each time-like bond.

In its present form, \Eq{loops_with_bubbles} has a sign problem associated with space-like vacuum bubbles on each time slice.
Particularly, the path-integral is weighted negatively whenever $\sum_\sigma \calF(c_\sigma)$ is odd.
One may eliminate the sign problem by explicitly summing up all possible fermion bubble contributions.
To see how this works, first note that the configurations $c\in C$ fall into classes identified by the path taken by closed-loops that wrap around the time direction; these classes are enumerated by the configurations $c\in C^*$, which involve no space-like fermion bubbles.
An example of two configurations that belong to the same class are provided in \Fig{config} (left, center), along with their unique bubble-free class-equivalent configuration (right).

Let me now carry out the sum over all configurations belonging to the same class.
First, note that for every configuration $c\in C$, we may identify a set of maximal one-dimensional space-like domains $\calD(c)$ associated with the sites left unvisited by closed loops, as demonstrated by the collection of shaded regions in \Fig{config} (right).
Note that every configuration belonging to the same class as $c$ shares a common set of domains $\calD$.
Next, consider a single domain $d\in\calD(c)$ belonging to a fixed time slice and of linear spatial extent $n=\calL(d)$, where $0\le n \le N_s$.
Since each domain is disjoint from all the others on every configuration belonging to the same class, we may assign an independent partition function $z_n(m)$ to each of these domains.
Specifically, I define $z_n(m)$ as a sum over the weights of all possible space-like fermion bubbles within a domain of length $n$.
With such a definition, one may show that the total sum over all configurations belonging to the same class as $c\in C^*$ will have a weight proportional to product of the partition functions $z_{\calL(d)}(m)$ associated with each of the domains $d\in\calD(c)$.

\begin{figure}
\includegraphics[width=\textwidth]{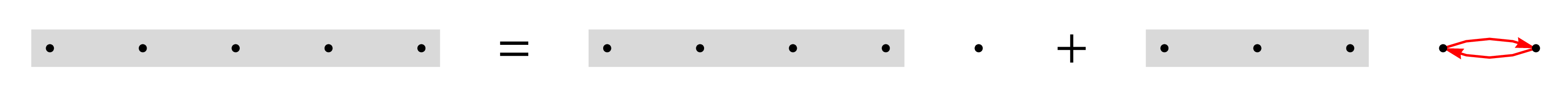}
\caption{%
\label{fig:recursion}%
Diagrammatic representation for the recursion relation defined in \Eq{recursion} for $z_5(m)$.
}
\end{figure}

I evaluate $z_n(m)$ explicitly for a given $n$ by noting certain algebraic properties of the one-dimensional partition function.
Particularly, the partition function satisfies the recursion relation:
\begin{eqnarray}
z_n(m) = z_{n-1}(m) \left(1+\frac{1}{m} \right) - \frac{1}{(2m)^2} z_{n-2}(m) \ ,\qquad 
z_0(m) = 1 \ ,\qquad
z_1(m) = 1+\frac{1}{m}\ ,
\label{eq:recursion}
\end{eqnarray}
which may be represented diagrammatically as in \Fig{recursion}.
Note that the negative sign appearing in the second term of the recursion relation arises from the presence of a space-like fermion bubble loop.
This recursion relation may be solved analytically and yields:
\begin{eqnarray}
z_n(m) = \frac{1}{2^{n+1}\sqrt{1+2/m}} \left[ \left(1 + \frac{1}{m} + \sqrt{1+\frac{2}{m}}  \right)^{n+1} -  \left(1 + \frac{1}{m} - \sqrt{1+\frac{2}{m}}  \right)^{n+1}  \right]\ ;
\end{eqnarray}
one may easily confirm that this solution is positive for all $m>0$ and any $n\ge0$. \footnote{This may also be seen by noting that $z_n(m) = \det \left[1 - \nabla^2/(2m) \right]$, for a Laplacian defined on a spatial lattice of length $n$ with open boundary conditions; the eigenvalues of the Laplacian are always real and positive.}.

Carrying out the summation over all configurations belonging to the same class as $c\in C^*$, I arrive at the final expression for the partition function:
\begin{eqnarray}
Z(\mu) = \sum_{ \{ c_\sigma \} \in C^* } \left[ \prod_\sigma  \left( \prod_{d\in\calD(c_\sigma)} z_{\calL(d)}(m_\sigma) \right) \left(\frac{1}{2m_\sigma} \right)^{\calB_s(c_\sigma)}  e^{\mu_\sigma \calB_\tau(c_\sigma)} \right] \left(1 + g \right)^{\calB_\tau( \cap_\sigma c_\sigma) } \ ,
\label{eq:final_result}
\end{eqnarray}
where the sum over configurations is now constrained to the set of all possible self-avoiding fermion loops that wrap around the time direction unaccompanied by space-like fermion bubbles.
This result is free of sign problems for any $m_\sigma>0$, $\mu_\sigma$ and $g \ge-1$, and is suitable for numerical simulation using a local loop-updating scheme or the ``worm algorithm'' \cite{2001PhRvL..87p0601P},  both at zero and finite temperatures.
The canonical ensemble partition function associated with a fixed set of charges $Q = \{Q_a, Q_b, Q_c, Q_d\}$ may be obtained via the Fourier transform of the grand-canonical partition function evaluated at imaginary chemical potential:
\begin{eqnarray}
\hat Z(Q) = \int_{-\pi}^\pi \left( \prod_\sigma \frac{d\theta_\sigma}{2\pi} \right) Z\left( \frac{i \theta}{N_\tau} \right) e^{-i Q_\sigma \theta_\sigma}.
\end{eqnarray}
Under this projection, the configurations that contribute to the canonical partition function satisfy the fixed winding number constraint $Q_\sigma = \calB_\tau(c_\sigma)/N_\tau$ and belong to a subset of $C^*$; an explicit expression for $\hat Z(Q)$ can be obtained from \Eq{final_result} by simply replacing:
\begin{eqnarray}
e^{\mu_\sigma \calB_\tau(c_\sigma)} \to \prod_\sigma \delta_{Q_\sigma, \calB_\tau(c_\sigma)/N_\tau}\ .
\end{eqnarray}

\section{External potentials}

One may introduce a general space-time and species-dependent external potential $v(x)$ in the continuum theory by replacing: $\mu \to \mu + v(x)$ in \Eq{continuum_action}.
On the lattice , this corresponds to a modification of the discretization rules:
\begin{eqnarray}
(\partial_\tau \psi - \mu\psi + v\psi)_\bfn &=& \frac{1}{b_\tau} \left( \psi_\bfn - e^{b_\tau (\mu - v_{\tilde\bfn} ) }  \psi_{\bfn-{\bfe_0}} \right) \ , \cr
(\psi^\dagger \psi)_\bfn            &=& \psi_\bfn^\dagger  e^{b_\tau ( \mu -v_{\tilde \bfn }) }  \psi_{\bfn-{\bfe_0}} \ ,
\end{eqnarray}
where $v_{\tilde\bfn} = \frac{1}{2} \left( v_\bfn + v_{\bfn-\bfe_0} \right)$, and $\tilde\bfn$ correspond to the half-lattice spacing locations $\tilde\bfn \equiv \bfn-\frac{1}{2} \bfe_0$ for each $\bfn\in\Lambda$.
As was the case with the chemical potential, an external potential is associated with the time-like bonds of the lattice and in a way may be viewed as a spacetime-dependent chemical potential.
At first glance, this result may seem rather strange, since in the naive discretization, an external potential would be associated with sites, rather than time-like bonds.
In fact, either choice is acceptable and should at most lead to different time discretization errors.
Here, I choose the point-split discretization as it leads to a rather simple expression for the path-integral in the loop representation.

In terms of a hopping parameter expansion, the introduction of an external potential suppresses fermion propagation in time at large potential energies.
Hence, if and when fermion loops are present, they will most likely be localized about the minimum of the potential, to the extent that Pauli-exclusion will allow.
In the loop representation, \Eq{final_result} is modified in an intuitive way, by replacing:
\begin{eqnarray}
e^{\mu_\sigma \calB_\tau(c_\sigma)} \to   e^{\mu_\sigma \calB_\tau(c_\sigma) - \calV_\sigma(c_\sigma)}\ ,
\end{eqnarray}
where
\begin{eqnarray}
\calV_\sigma(c) = \frac{1}{2} \sum_{\tilde\bfn \in c} \left( v_{\sigma,\tilde\bfn+\frac{1}{2} \bfe_0} +  v_{\sigma,\tilde\bfn-\frac{1}{2} \bfe_0} \right) \ ,
\end{eqnarray}
and the sum on $\tilde\bfn \in c$ is over the half-lattice sites that lie on the time-like bonds of the configuration $c$.

Provided that the external potential is confining, we may explicitly take the infinite volume limit without worrying about fermions wondering off to infinity.
The reason we may do this is that the volume-dependence only appears in $z_{\calL(d)}(m)$ for the domains $d$ that touch the boundaries of the lattice.
In any local updating scheme, only the ratios $R_n^\pm(m) = z_{n \pm 1}(m) /z_n(m)$ matter; for the boundary domains, taking $N_s\to\infty$ corresponds to taking the $n\to\infty$ limit of this ratio and yields:
\begin{eqnarray}
R_\infty^\pm (m) = \left[ \frac{1}{2} \left( 1+\frac{1}{m} + \sqrt{1+\frac{2}{m}}\right)  \right]^{\pm 1} \ .
\end{eqnarray}

\section{Observables}

Short of explicitly introducing sources and sinks for the fields $\psi^\dagger_\bfn$ and $\psi_\bfn$, there are several observables which may easily be calculated using the above path-integral representation at finite temperature and density.
Examples include derivatives of the partition function with respect to masses, potentials and the coupling, which yield expectation values for the kinetic energy operators ($T_\sigma$), potential energy operators ($V_\sigma$), and interaction energy operator ($I$), respectively.
More concretely, the expectation value of the kinetic energy operator is given by:
\begin{eqnarray}
T_\sigma + \hat T_\sigma = -\frac{1}{N_\tau} \frac{\partial \log Z(\mu)}{\partial \log(1/m_\sigma)}
                         =  -\frac{1}{N_\tau} \left\langle \sum_{d\in \calD(c_\sigma)} \frac{\partial  \log z_{\calL(d)}(m_\sigma)  }{\partial\log (1/m_\sigma)}  +  \calB_s(c_\sigma)  \right\rangle\ ,
\end{eqnarray}
where 
\begin{eqnarray}
\hat T_\sigma =  - \frac{ \partial \log z_{N_s}(m_\sigma) }{ \partial\log (1/m_\sigma) } 
\end{eqnarray}
is a zero temperature and density contribution, which must be subtracted off.
The expectation value of the potential and interaction operators are given by:
\begin{eqnarray}
V_{\sigma} = -\frac{1}{N_\tau}  \sum_{\tilde \bfn} \frac{\partial \log Z(\mu) }{\partial \log v_{\sigma,\tilde \bfn} } = \frac{1}{N_\tau} \left\langle \calV_\sigma(c_\sigma) \right\rangle
\end{eqnarray}
and 
\begin{eqnarray}
I = \frac{1}{N_\tau} \frac{\partial \log Z(\mu)}{\partial \log g } = \frac{1}{N_\tau}  \frac{g}{1+g} \left\langle \calB_\tau( \cap_\sigma c_\sigma) \right\rangle  \ ,
\end{eqnarray}
respectively.
In addition to these, charges and susceptibilities associated with each species are obtained from the expressions:
\begin{eqnarray}
Q_\sigma = \frac{1}{N_\tau} \frac{\partial }{\partial \mu_\sigma } \log Z(\mu) =  \frac{1}{N_\tau} \langle \calB_\tau(c_\sigma) \rangle\ ,
\label{eq:charge}
\end{eqnarray}
and
\begin{eqnarray}
 \chi_{\sigma\rho} =  \frac{1}{N_\tau} \frac{\partial^2 }{\partial \mu_\sigma \partial \mu_\rho } \log Z(\mu)
                   = \frac{1}{N_\tau} \left[\langle \calB_\tau(c_\sigma) \calB_\tau(c_\rho)  \rangle -\langle \calB_\tau(c_\sigma) \rangle \langle \calB_\tau(c_\rho) \rangle  \right] \ ;
\end{eqnarray}
these just measure the average winding numbers associated with the close loop configurations for each species, and their fluctuations.
Combining these results, the total free-energy, $F(\mu) = -\log{Z(\mu)}/N_\tau$, of the system in the grand-canonical ensemble is given by:
\begin{eqnarray}
F(\mu) = \sum_\sigma \left( T_\sigma  + V_\sigma -\mu_\sigma Q_\sigma \right) + I 
\end{eqnarray}
at zero temperature.
In the canonical ensemble, where the charges $Q_\sigma$ are held fixed and the corresponding susceptibilities are zero (one should use an updating scheme designed to maintain such constraints), the expectation value of the total energy of the system is given by:
\begin{eqnarray}
E(Q) = \sum_\sigma \left( T_\sigma  + V_\sigma \right) + I \ .
\end{eqnarray}

In the $SU(4)$ symmetric limit, a quantity which is closely related to the interaction energy, known as the integrated contact density, is given by  $\calC = (mg)^2 I /g $.
This quantity appears in a set of universal relations, including the generalized Virial theorem:
\begin{eqnarray}
E = 2 \sum_\sigma V_\sigma - \frac{\calC}{8\pi m a}\ ,
\end{eqnarray}
valid in the presence of a time- and species-independent harmonic potential $v_{\tilde \bfn} = \frac{\kappa}{2} (n_s-\bar n_s)^2$ centered about $\bar n_s$ with spring constant $\kappa$, and the adiabatic relation:
\begin{eqnarray}
\frac{\partial E}{\partial (-1/a)} = \frac{\calC}{4\pi m}\ .
\end{eqnarray}
These relations may tested straight-forwardly in a numerical simulation using the loop representation for the partition function.

Finally, density correlations may also be computed in the loop representation by considering functional derivatives of the partition function with respect to $v_{\tilde \bfn}$, the exponential of which couples to the density operator $\rho_{\tilde \bfn} = \psi^\dagger_{\bfn} e^{\mu} \psi_{\bfn-\bfe_0} $.
Generally, one may consider $N$-point functions of the form:
\begin{eqnarray}
\Gamma_{\sigma_1,\ldots,\sigma_N}(\tilde\bfn_1,\ldots,\tilde\bfn_N) = \left\langle  \prod_{j=1}^N \rho_{\sigma_j,\tilde \bfn_j} \right\rangle = \left( \prod_{j=1}^N \frac{ \partial }{\partial v_{\sigma_j, \tilde \bfn_j }} \right)  \log Z(\mu)\ ;
\label{eq:density_correlations}
\end{eqnarray}
if I define the local density field:
\begin{eqnarray}
\phi_{\tilde\bfn}(c) = \sum_{\tilde\bfm\in c} \delta_{\tilde\bfm,\tilde\bfn}\ ,
\end{eqnarray}
then the number density of the system, $n_\sigma \equiv \Gamma_{\sigma}(\tilde \bfn)$, is just given by the 1-point correlator:
\begin{eqnarray}
\Gamma_{\sigma}(\tilde \bfn)  = \left\langle \phi_{\tilde\bfn}(c_\sigma) \right\rangle\ .
\end{eqnarray}
Summing $\Gamma_{\sigma}(\tilde \bfn) $ over all space at an arbitrary time slice yields \Eq{charge} for the expectation value of the charge of the system.
Similarly the 2-point function (i.e., the density-density correlator) is given by:
\begin{eqnarray}
\Gamma_{\sigma_1,\sigma_2}(\tilde\bfn_1,\tilde \bfn_2) = 
\langle \phi_{\tilde\bfn_1}(c_{\sigma_1}) \phi_{\tilde\bfn_2}(c_{\sigma_2}) \rangle - \langle \phi_{\tilde\bfn_1}(c_{\sigma_1}) \rangle \langle \phi_{\tilde\bfn_2}(c_{\sigma_2}) \rangle\ .
\end{eqnarray}
All other density correlators may be derived in terms of $\phi_{\tilde\bfn}(c)$ from \Eq{density_correlations} in a straight-forward fashion.

\section{Mean-field Analysis}

The continuum limit for this theory corresponds to the simultaneous limits: $a\to0$ and $n_\sigma\to0$, where $a$ and $n_\sigma$ are measured in units of the lattice spacing, while keeping the dimensionless quantity $\eta = (p_F a)^{-1}$ held fixed.
Here, I investigate whether there exists such a regime in ($g,\mu_\sigma,m_\sigma$) parameter space, where a continuum limit may be defined for any $\eta$.
I carry out the analysis within mean-field theory, deriving self-consistent equations for the number densities, and an upper bound $f_{MF}(\mu)$ on the free-energy density $f(\mu) = F(\mu)/N_s$, valid at finite temperature, for an arbitrary spin and mass imbalance, and for any coupling $g>-1$.
I numerically solve the self-consistent equations at zero temperature and study the phases of the model in the $SU(4)$ limit of equal masses and chemical potentials for each species.

On general grounds, from the positivity of \Eq{final_result}, one may show that the free-energy density of the interacting four-component Fermi gas is bounded by \cite{springerlink:10.1007/BF02418571}:
\begin{eqnarray}
f_{MF}(\mu) = f_{\lambda}(\mu) + \frac{1}{N_\tau N_s} \left\langle S(\mu) - S_\lambda(\mu) \right\rangle_{\lambda}\ ,
\label{eq:mean-field}
\end{eqnarray}
where $f_\lambda(\mu)$ is the free-energy density of some arbitrary reference system described by the action $S_\lambda(\mu)$, the expectation value of the difference in actions is taken with respect to the reference system, and $\lambda$ is some variational parameter(s).
The tightest bound on $f(\mu)$ is achieved by minimizing $f_{MF}(\mu)$ with respect to $\lambda$.
From \Eq{final_result}, the action associated with the partition function for the interacting Fermi gas in the loop representation is given by:
\begin{eqnarray}
S(\mu) = \sum_\sigma S_{FG}(\mu_\sigma) - \log(1+g) \calB_\tau(\cap_\sigma c_\sigma)\ ,
\end{eqnarray}
where $S_{FG}(\mu)$ is the action for a non-interacting single component Fermi gas at finite chemical potential.
For the purpose of this calculation, I will choose as my reference action:
\begin{eqnarray}
S_\lambda(\mu) = \sum_\sigma S_{FG}(\mu_\sigma ) - \sum_\sigma \lambda_\sigma \calB_\tau(c_\sigma) \ ;
\label{eq:reference_pf}
\end{eqnarray}
note from the form of \Eq{reference_pf}, that $S_\lambda(\mu) = \sum_\sigma S_{FG}(\mu_\sigma + \lambda_\sigma)$, and therefore it follows that $f_\lambda(\mu) =  \sum_\sigma f_{FG}(\mu_\sigma+\lambda_\sigma)$.
Evaluation of \Eq{mean-field} with respect to the reference action yields
\begin{eqnarray}
f_{MF}(\mu) = \sum_\sigma f_{FG}(\mu_\sigma+\lambda_\sigma) - \log(1+g) \prod_\sigma n_{FG}(\mu_\sigma+\lambda_\sigma) + \sum_\sigma  \lambda_\sigma n_{FG}(\mu_\sigma+\lambda_\sigma) \  ,
\label{eq:mf_fg}
\end{eqnarray}
where
\begin{eqnarray}
f_{FG}(\mu) = \hat f_{FG} -\frac{1}{\pi} \int_0^\pi dp\, \left[ \frac{1}{N_\tau}\log\left( 1 + e^{-N_\tau (\log  \xi_p(m) - \mu) } \right) \right] \ ,
\end{eqnarray}
and
\begin{eqnarray}
\hat f_{FG} = -\frac{1}{\pi} \int_0^\pi dp \log{\xi_p(m)}\ ;
\end{eqnarray}
note that $\hat f_{FG}$ is an irrelevant $\mu$-independent shift in the free-energy density.
To arrive at \eq{mf_fg}, I have used the relation:
\begin{eqnarray}
\calB_\tau(\cap_\sigma c_\sigma) = \sum_{\tilde\bfn} \prod_\sigma \phi_{\tilde\bfn}(c_\sigma)\ ,
\end{eqnarray}
and the fact that an expectation value of the summand on the right-hand-side with respect to the reference system factorizes into a product of number densities associated with each species, each of which is independent of $\tilde\bfn$ at infinite volume.

The mean-field free-energy density is minimized with respect to $\lambda$ when
\begin{eqnarray}
\lambda_\sigma  = \log(1+g) \prod_{\sigma^\prime \neq \sigma} n_{FG}(\mu_{\sigma^\prime}+\lambda_{\sigma^\prime})\ ,
\label{eq:extremization}
\end{eqnarray}
where the free-gas number density is given by
\begin{eqnarray}
n_{FG}(\mu) = - \frac{\partial}{\partial \mu_\sigma} f_{FG}(\mu)\ ;
\end{eqnarray}
the mean-field density is in turn related to the free-gas number density by
\begin{eqnarray}
n_{MF}(\mu) = \sum_\sigma n_{FG}(\mu_\sigma + \lambda_\sigma)\ .
\end{eqnarray}
Specializing to the case of zero temperature,\footnote{Here, I make use of the identity: $\lim_{\beta\to\infty}  \log(1+x^\beta)/\beta = \theta(x-1) \log(x)$.} the free-energy density for a single species becomes:
\begin{eqnarray}
f_{FG}(\mu) = \hat f_{FG} +  \calE_{FG}(\mu) - \mu n_{FG}(\mu)\ ,
\end{eqnarray}
where
\begin{eqnarray}
\calE_{FG}(\mu) = \frac{1}{\pi} \int_0^{p_F(\mu)} dp \log\xi_p(m)\ ,\qquad n_{FG}(\mu) = \frac{1}{\pi} p_F(\mu)\ ,
\end{eqnarray}
are the corresponding energy and number densities defined in terms of the Fermi momentum:
\begin{eqnarray}
p_F(\mu) = 
\left\{ \begin{array}{ll}
2 \arcsin\sqrt{\frac{m}{2}\left( e^{\mu}-1\right)}  & \mu < \mu_0 \\
\pi  &   \mu > \mu_0 \ ,
\end{array} \right.
\end{eqnarray}
and $\mu_0 = \log\left( 1+2/m\right)$ is the free-gas chemical potential at which the lattice becomes saturated by fermions.

\begin{figure}
\includegraphics[width=0.43\textwidth]{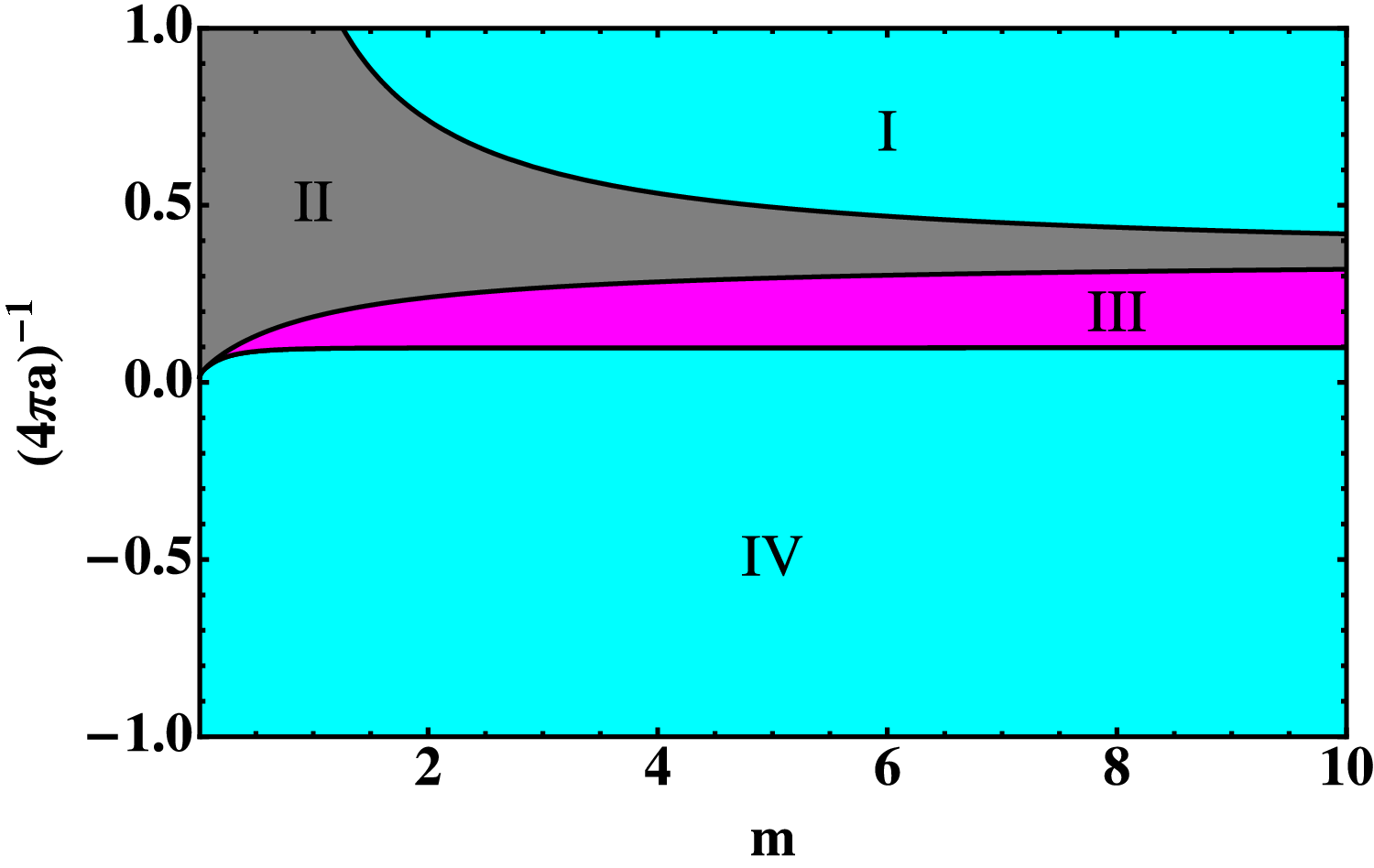}
\hspace{0.5in}
\includegraphics[width=0.4\textwidth]{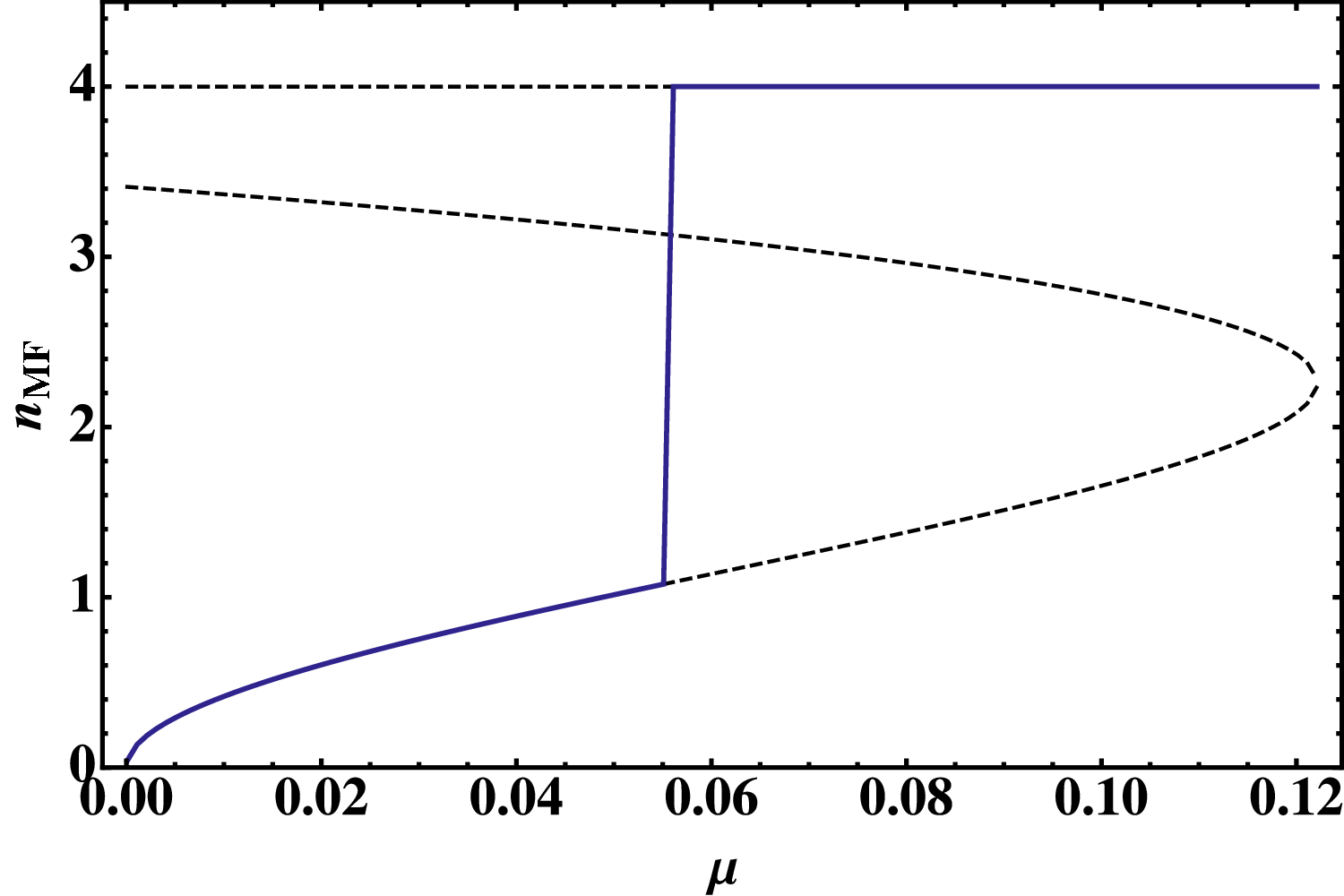}
\caption{%
\label{fig:phases}%
Left: number density phase diagram in the $m-a^{-1}$ plane at $\mu=0$, computed within mean-field theory.
(I) and (IV) correspond to regions of vanishing number density, (III) corresponds to a region in which the lattice is fully saturated by fermions, and (II) is a region that is inaccessible by a mean-field treatment.
A continuum limit can only exists in region (IV), where both the inverse number density (i.e., interparticle spacing) and scattering length can be taken large compared to the lattice spacing.
Lines of separation between each of these regions are defined in the text.
Right: number density as a function of chemical potential for a fixed fermion mass $m=5$ and $a=\infty$ (i.e., starting from a point in region IV at $\mu=0$).
Dashed lines indicate extrema of the free-energy density, whereas the solid line indicates the solution to \Eq{extremization} which minimized the free-energy density globally.
The figure exhibits a first order phase transition from a region of small number density to a region in which the lattice is fully saturated by fermions.
The continuum limit corresponds to the limit of zero density in lattice units, or $\mu\to0$.
}
\end{figure}

To simplify the analysis further, let me now consider the $SU(4)$ limit in which the Fermi gas is unpolarized, and the mass of each species is equal.
I begin by considering the number density phase diagram in the $m-a^{-1}$ plane at $\mu=0$, represented by \Fig{phases} (left).
Within mean-field theory, this phase diagram is characterized by four regions, labeled (I), (II), (III) and (IV), separated by curves corresponding to the couplings: $g_{I-II} = -1$, $g_{II-III} = \infty$, and $g_{III-IV} = e^{-4 \hat f_{FG}} - 1$.
Region (I) corresponds to repulsive interaction strengths ranging from $g_{I-II}<g<0$, and has vanishing number density within mean-field theory.
Region (II) corresponds to a regime of strong repulsive interaction strengths ranging from $-g_{II-III} < g < g_{I-II}$, and is inaccessible within a mean-field theory treatment due to the non-positivity of \Eq{final_result}.
Region (III) corresponds to strong attractive interaction strengths ranging from $g_{III-IV} < g < g_{II-III}$; in this regime, the number density is four, and corresponds to a lattice that is fully saturated by fermions.
Finally region (IV) corresponds to attractive interaction strengths ranging from $0<g<g_{III-IV}$; in this regime the number density vanishes, allowing for the existence of continuum physics.

Extending \Fig{phases} (left) into a third dimension labeled by the chemical potential $\mu$, the critical line defined by the coupling $g_{III-IV}$ becomes a critical surface separating a domain which is continuously connected to the zero density region (IV) from a domain which is continuously connected to the saturated region (III).
In \Fig{phases} (right), I plot the number density as a function of chemical potential for a fixed mass $m=5$ and $a=\infty$.
Zero chemical potential on this plot corresponds to a point in region (IV) in \Fig{phases} (left); as the chemical potential is increased, there exists a critical value at which the density changes discontinuously and one enters into the phase that is continuously connected to region (III).
Below this critical chemical potential, a region exists in which the density may be taken arbitrarily small and the scattering length arbitrarily large such that $\eta$ may take on a range of finite values, suggesting a continuum limit exists within the mean-field analysis.
Note that the continuum limit places no restriction on the mass parameter; the mass $m$ (or more generally, the reduced mass $ \bar m = \left( \sum_\sigma 1/m_\sigma \right)^{-1}$ in the mass-imbalanced case) may instead be regarded an anisotropy factor for the temporal and spatial lattice spacings.

\section{Discussion}
I have demonstrated that the partition function for the four-component universal Fermi gas proposed in \cite{Nishida:2009pg} may be formulated on the lattice in a way that avoids the fermion sign problem at arbitrary chemical potentials, mass imbalances and sign of the four-body interaction strength.
The formulation involved a path-integral over self-avoiding fermion loops, and is suitable for numerical simulation using standard local updating schemes.
Furthermore, I have derived expressions for the expectation values of all relevant observables, including the kinetic, potential and interaction energies, moments of the charge distributions, as well as a general expression for N-point correlation functions involving the density operator.
These observables may be directly related to interesting nonperturbative quantities in the BEC and cross-over regimes, as discussed in \cite{Nishida:2009pg}; full-scale numerical simulations to compute these quantities will be the subject of a future study.

Positivity of the path-integral in the loop representation permits a mean-field analysis of the lattice theory.
In the $SU(4)$ limit, I have demonstrated within this analysis that there exists a phase in which both $1/a$ and $n$ may be taken to zero (in lattice units) while achieving a range of finite values for $\eta$, thus allowing one to define a continuum limit for the lattice theory.
It would be interesting to extend this mean-field analysis to the population- and mass-imbalanced interacting Fermi gas, for which the path-integral representation retains its positivity property; ultimately, a nonperturbative study of the $(g,\mu_\sigma,m_\sigma)$ phase diagram is desirable as well.

The formalism presented here may be generalized to include two- and three-body contact interactions in a straightforward manner.
Extending the approach to higher dimensions, while straight-forward, yields an action with a sign problem, however.
The reason is that in higher dimensions, a single fermion loop may wrap more than once around the lattice in the time direction without crossing itself.
Generally a negative sign accompanies every fermion loop, and an additional negative sign accompanies every instance that the fermion wraps around the lattice in time, due to the choice of anti-periodic boundary conditions.
A relative negative sign in the weight will therefore always persist whenever the winding number minus the total number of fermion loops wrapping in time is odd.

Finally, it was observed that in one spatial dimension, one may regard a nonrelativistic Fermi gas as a Bose gas with an impenetrable hard core \cite{1960JMP.....1..516G}.
In light of this, it may seem unsurprising that the Fermi system may be formulated in a way that avoids the sign problem if the corresponding bosonic system is also free of sign problems.
It is known, however, that the conventional path-integral formulation for bosons does in fact have a sign problem at finite density.
Interestingly, the sign problem for bosons at finite density may be avoided by resorting to techniques similar to those discussed here \cite{PhysRevD.75.065012}.

\begin{acknowledgments}
I would like to thank J.-W. Chen, D. B. Kaplan, Y. Nishida, D. T. Son, and H. Suzuki for interesting and helpful discussions.
I am grateful to the national Institute for Nuclear Theory, where portions of this work took place.
\end{acknowledgments}

\bibliography{unitary_loop}

\end{document}